\documentclass[useAMS,usenatbib,twocolumn]{mn2e}

\usepackage{graphicx}
\usepackage{placeins}
\usepackage{flushend}
\usepackage{cuted}
\usepackage{widetext}
\usepackage{bm}
\usepackage{natbib}
\bibpunct{(}{)}{;}{a}{}{,}
\usepackage{amssymb}
\usepackage{amsmath}
\usepackage{times}
\usepackage[]{eufrak}
\usepackage{color}
\usepackage{textcomp}
\usepackage{txfonts}

\newcommand{\be}{\begin{equation}}
\newcommand{\ee}{\end{equation}}
\newcommand{\bdm}{\begin{displaymath}}
\newcommand{\edm}{\end{displaymath}}
\newcommand{\bea}{\begin{eqnarray}}
\newcommand{\eea}{\end{eqnarray}}

\title[Explaining radio emission of magnetars]
{Explaining radio emission of magnetars via
 rotating and
oscillating magnetospheres of neutron stars}

\author[V.~S.~Morozova, B.~J.~Ahmedov and O.~Zanotti]
        {Viktoriya~S.~Morozova$^{(1,\;2,\;3)}$,
          Bobomurat~J.~Ahmedov$^{(1,\;2,\;3)}$  and
        Olindo~Zanotti$^{(4)}$
                                                                \\
 $^{(1)}$Institute of Nuclear Physics,
        Ulughbek, Tashkent 100214, Uzbekistan                   \\
        $^{(2)}$Ulugh Beg Astronomical Institute,
        Astronomicheskaya 33, Tashkent 100052, Uzbekistan       \\
        $^{(3)}$The Abdus Salam International Centre
for Theoretical Physics, 34151 Trieste, Italy\\
$^{(4)}$Max-Planck-Institut f$\ddot{u}$r
        Gravitationsphysik, Albert-Einstein-Institut, 14476
        Golm, Germany\\
\\}
\pagerange{\pageref{firstpage}--\pageref{lastpage}}

\pubyear{2011}

\begin{document}

\maketitle

\label{firstpage}

\begin{abstract}
We investigate the conditions for radio emission in rotating and
oscillating magnetars, by focusing on the main physical processes
determining the position of their death-lines in the $P-\dot{P}$
diagram, i.e. of those lines that separate the regions where the
neutron star may be radio-loud or radio-quiet. After using the
general relativistic expression for the electromagnetic scalar
potential in the magnetar magnetosphere, we find that larger
compactness parameters of the star as well as larger inclination
angles between the rotation axis and the magnetic moment produce
death-lines well above the majority of known magnetars. This is
consistent with the observational evidence of no regular radio
emission from the magnetars in the frequency range typical for the
ordinary pulsars. On the contrary, when oscillations of the magnetar
are taken into account, the death-lines shift downward and the
conditions necessary for the generation of radio emission in the
magnetosphere are met. Present observations showing a close
connection between the burst activity of magnetars and the
generation of the radio emission in the magnetar magnetosphere are
naturally accounted for within our interpretation.

\end{abstract}

\begin{keywords}
{MHD: pulsars --- general --- relativity --- oscillations --- magnetar
--- stars: neutron --- death line --- plasma magnetosphere}
\end{keywords}

\section{Introduction}

Magnetars are neutron stars with very strong magnetic fields, namely
$B_0\simeq10^{14}-10^{15} \mathrm{G}$ \citep{Duncan1992}, in
comparison with the magnetic field of ordinary pulsars,
$B_0\simeq10^{12} \mathrm{G}$. At the moment $21$ magnetars are
known\footnote{See the continuously updated on-line catalog at the
hurl http://www.physics.mcgill.ca/\texttildelow
pulsar/magnetar/main.html.}, 9 of which as soft gamma-ray repeaters
(SGRs) and 12 of which as  anomalous X-ray pulsars (AXPs). Magnetars
distinguish from ordinary pulsars also because they have larger
periods of rotation, $P\simeq5-10\ \mathrm{s}$, and they spin down
much faster, with typical $\dot{P}\simeq10^{-10}-10^{-12}$ in
contrast to $\dot{P}\simeq10^{-15}$ for ordinary pulsars\footnote{
$P$ is the period of the neutron star, while $\dot{P}$ is the time
derivative of the period.}. These numbers provide the huge magnetic
fields reported above, through the simple dipolar approximation
$B_0\simeq2(P\dot{P}_{-15})^{1/2}10^{12} \mathrm{G}$. In addition,
magnetars are characterized by persistent X-ray luminosities
$L_X\approx 10^{34}-10^{36}\rm{erg} \ s^{-1}$, no evidence of
Doppler shifts in the light curves and lack of any bright optical
companion. The canonical magnetar model, proposed by
\citet{Duncan1992}, is based on the consideration that, unlike the
case of radio pulsars, the global energetics of magnetars is
accounted for by magnetic energy, rather than rotational
energy~\citep{Kramer2008}. Recent reviews reporting both theoretical
modeling and observations of SGRs and AXPs can be found in
\cite{Woods2006} and \citet{Hurley2009}.

The activity of magnetars is observed in the form of bursts in X-ray
and $\gamma$-ray bands, while there is no periodic radio emission
from the majority of magnetars in the same range of frequencies of
ordinary pulsars. It was recently shown by \cite{Istomin2007}
(hereafter IS07) that the absence of radio emission from magnetars
is likely to be related to their slow rotation, which would also
explain the low energy of the primary particles, accelerated near
the surface of the star. IS07 has also investigated the physics
determining the slope and the position of the death-line for
magnetars, i.e. the line in the $P-\dot{P}$ diagram that separates
the regions where the neutron star may be radio-loud or
radio-quiet\footnote{ Additional information about the neutron star
death-line can be found in \cite{Ruderman1975}, \cite{Arons1979},
\cite{Chen1993}, \cite{Rudak1994}, \cite{Zhang2000},
\cite{Kantor2004}.}.

However, in the recent past two magnetars, namely XTE J1810|197 and
1E 1547.0-5408~\citep{Camilo2006,Camilo2007}, plus the candidate
source PSR J1622-4950~\citep{Levin2010}, were discovered that
present radio emission very similar to that encountered in ordinary
pulsars, but still with some peculiar features. It is interesting to
note that in both the radio-magnetars AXP J1810-197 and 1E
1547.0-5408, the radio emission was anticipated by a X-ray burst.
Moreover, recent observations of quasi-periodic oscillations in the
initially rising spike and decaying tail of the spectra of SGRs
\citep{Barat1983, Israel2005,
  Terasawa2005, Strohmayer2006} suggest that
neutrons stars in general, and magnetars in particular, may be
subject to some seismic events, so called glitches, producing
mechanical oscillations of the stellar crust~\citep{Schumaker1983,
McDermott1988, Duncan1998, Glampedakis2007, Timokhin2008,
Eichler2010, Colaiuda2011}, and providing the opportunity to infer
crucial information about the internal structure of these
objects~\citep{Levin2007, vanHoven2011}. Oscillations of the stellar
crust, as in the case of the earthquakes, may cause electromagnetic
events in the plasma magnetosphere of the neutron stars and have
influence on the parameters of this magnetosphere. Investigations of
the electrodynamics of the oscillating neutron star magnetosphere
have been performed by \cite{Timokhin2000}, \cite{Abdikamalov2009},
\cite{Ahmedov2009} and \cite{Morozova2010}. The burst character of
magnetars activity indicates that they may be subject to different
kind of perturbations (and therefore oscillations) with a
probability higher than that found in ordinary pulsars. In the
present paper we consider the influence of magnetar oscillations on
the conditions for the radio emission generation in the
magnetosphere of magnetars. In particular, we revisit the problem of
magnetars death-line, by taking into account the role both of
rotation and of toroidal oscillations in a relativistic framework.
Although our analysis follows the general logic presented in IS07,
for the electromagnetic scalar potential in the magnetosphere of the
neutron stars we adopt the more consistent expressions reported by
\citet{Muslimov1992} and \cite{Morozova2010}. We show that, by
virtue of this modification, the lack of radio-emission from
magnetars, at least in the radio-band typical for the ordinary
pulsars, can be naturally explained. Moreover, we also show that, as
in the case of ordinary pulsars~\citep{Morozova2010}, the
oscillations make the death-line shift down, allowing for some
magnetars to become radio-loud.

The plan of the paper is the following. In Section~\ref{sect1} we
discuss the properties of the death-line for magnetars, while
introducing some basic concepts. Subsection~\ref{subsect11}, in
particular, is devoted to the aligned magnetars and it considers the
dependence of the death-line position on the compactness parameter
of the magnetar. while in Subsection~\ref{subsect12} we investigate
the dependence of the death-line position on the inclination angle
of the magnetar. Section~\ref{deathline}, on the other hand, is
devoted to the case of rotating as well as oscillating magnetars.
Finally, Section~\ref{conclusions} contains the conclusions of our
work.

In the rest of the paper we adopt a system of units for which
$\hbar=\lambdabar=c=1$, where $\hbar$ is the Planck constant,
$\lambdabar$ is the Compton wavelength of the electron and $c$ is
the speed of light. The final estimations for the magnetic field are
given in Gauss.

%-----------------------------------------------------
\section{Relativistic death-line for magnetars}
\label{sect1}

\subsection{Basic concepts}

IS07 investigated the distribution function of electrons and
positrons in the magnetosphere of magnetars and they showed that the
possible Lorentz factor of the particles in electron-positron plasma
is restricted by the boundary values $\gamma_{min}$ and
$\gamma_{max}$. The maximum value $\gamma_{max}$ is determined by
the energy $k$ of the photons producing the electron-positron pairs
in the neutron star magnetosphere. These photons are emitted by the
primary particles, pulled out from the surface of the neutron star
and accelerated to ultra-relativistic velocities very close to the
star surface due to the presence of the unscreened component of the
electric field parallel to the magnetic field. The characteristic
energy of these photons, called curvature photons, is given by
$k=3\gamma_0^3/2\rho$, where $\gamma_0$ is the Lorentz factor of the
primary accelerated particles and $\rho$ is the radius of curvature
of the magnetic field lines in the point of emission. The creation
of the electron-positron pair by the single photon moving in the
magnetic field of the star is possible if the angle $\delta$ between
the trajectory of the photon and the magnetic field lines reaches
some threshold value $\delta_t=2/k$. As it was shown in
\cite{Beskin1993} the energy of the photon generating the pair is
distributed between the electron and positron almost evenly and the
Lorentz factor of created particles is determined by the expression
$\gamma=1/\delta$ (see also \cite{Beskin2010}). Thus, one may find
the maximum value of the Lorentz factor of the produced particles as
$\gamma_{max}=1/\delta_t\approx3\gamma_0^3/4\rho$, which, obviously,
decreases with increasing radius of curvature of the magnetic field
lines and depends on the energy of the primary accelerated
particles.

The minimum value of the Lorentz factor of particles in the
electron-positron plasma magnetosphere of the neutron star
$\gamma_{min}$ is determined by the geometry of the magnetic field
of the neutron star and increases with the increasing radius of
curvature of magnetic field lines. IS07 found
$\gamma_{min}=\rho/z_0$, where $z_0$ is the distance of the photon
emission point from the center of the star along the dipole axis.
The key point of the discussion, which allows to determine the slope
of the death line in the $P-\dot{P}$ diagram, is that, by equating
$\gamma_{min}$ and $\gamma_{max}$, one can get the condition for the
minimum value of the magnetic field for which the production of the
electron-positron pairs in the magnetosphere of the neutron star is
still possible.

IS07 have also shown that the condition $\gamma_{min}=\gamma_{max}$
is equivalent to the condition $l_{f\ min}=R_s$, where $l_f$ is the
mean free path of the photons emitted by primary particles near the
stellar surface and $R_s$ is the stellar radius. After using this
condition, IS07 computed the minimum value of the magnetic field
allowing for effective electron-positron plasma production within
the magnetosphere of a rotating magnetar. In particular, the slope
of the death-line of ordinary pulsars, with surface magnetic fields
of the order of $10^{12} \mathrm{G}$, turns out to be $11/4$, while
that of magnetars,  with surface magnetic fields of the order of
$10^{15} \mathrm{G}$, is $11/3$. This is due to the different
absorption coefficients for the curvature photons in the case of
relatively weak magnetic fields, namely when $B\gtrsim B_c$, and in
the case of the huge magnetic fields of magnetars\footnote{The
critical magnetic field is defined
  as $B_c=m^2c^3/e\hbar\approx  4.414\times10^{13}
  \mathrm{G}$,  where $m$ is the electron mass and $e$ is the electron
charge.}, namely when $B>>B_c$.

The minimum possible mean free path of curvature photons can be
found using the relation $l_f=\rho/\gamma_{max}$ (see IS07) with
$\gamma_{max}=3\gamma_0^3/4\rho$, taking into account that
$\gamma_0$ is determined by the the scalar potential accelerating
the first generation of particles, i.e. $\gamma_0=|\Psi(\theta,\phi)|$.

Here we assume that the structure of the magnetar magnetosphere has
the same qualitative features of the magnetosphere of ordinary
pulsars. In particular, we assume that the magnetic field of the
magnetar has a dipolar structure, with a region of closed field
lines co-rotating with the star, and, in addition, a region of so
called open magnetic field lines, escaping to infinity through the
surface of the light cylinder. Moreover, at the surface of the star
the open magnetic field lines form  a region, known as the {\em
polar cap region}, which is very relevant in the context of particle
acceleration mechanisms within neutron star magnetospheres. Unlike
IS07, who adopted a Newtonian approximation for the  electromagnetic
scalar potential in the vicinity of the polar cap region, in our
analysis we have used the consistent relativistic expression
provided by \citet{Muslimov1992}, which is valid at angular
distances $\Theta_0<<\eta-1<<R_c/R_s$, i.e.

\begin{widetext}
\begin{equation}
\label{Mus_phi}
\Phi=\frac{1}{2}\Phi_0\kappa\Theta_0^2\left(1-\frac{1}{\eta^3}\right)(1-\xi^2)\cos\chi+\frac{3}{8}\Phi_0\Theta_0^3
H(1)\left(\frac{\Theta(\eta)H(\eta)}{\Theta_0
H(1)}-1\right)\xi(1-\xi^2)\sin\chi\cos\phi\ ,
\end{equation}
\end{widetext}

with
\begin{eqnarray}
H(\eta)&=&\frac{1}{\eta}\left(\varepsilon-\frac{\kappa}{\eta^2}\right)+
\nonumber \\
&&\left(1-\frac{3}{2}\frac{\varepsilon}{\eta}
+\frac{1}{2}\frac{\kappa}{\eta^3}\right)\left[f(\eta)\left(1-\frac{\varepsilon}{\eta}\right)\right]^{-1}\
, \nonumber \\
f(\eta)&=&-3\left(\frac{\eta}{\varepsilon}\right)^3\left[\ln\left(1-\frac{\varepsilon}{\eta}\right)+
\frac{\varepsilon}{\eta}\left(1+\frac{\varepsilon}{2\eta}\right)\right]\
,
\end{eqnarray}
where $\eta=r/R_s$ is the dimensionless radial coordinate,
$\Theta(\eta)$ is the polar angle of the last open magnetic field
line, given by
\begin{equation}
\label{theta}
\Theta(\eta)\cong\sin^{-1}\left\{\left[\eta\frac{f(1)}{f(\eta)}\right]^{1/2}\sin\Theta_0\right\}
\ ,
\end{equation}
$\Theta_0$ is the polar angle of the last open field
line at the surface of the star, given by\footnote{
Note that the small angles approximation is applicable
since the radius of the polar cap is of the order of a
few hundred meters,
while the typical value for the radius of the star is $10
\mathrm{km}$.}
\begin{equation}
\label{theta0} \Theta_0=\sin^{-1}\left(\frac{R_s}{R_c
f(1)}\right)^{1/2}\ ,
\end{equation}
while $\Omega$, $R_c=1/\Omega$, $\Phi_0=\Omega B_0 R_s^2$  are the
angular velocity of the neutron star rotation, the radius of the
light cylinder and the characteristic value of scalar potential
generated in the vicinity of the neutron star, respectively.
Finally, $\chi$ is the inclination angle between the angular
momentum of the neutron star and its magnetic moment,
$\varepsilon=2GM/R_s$ is the compactness parameter, $G$ is the
gravitational constant, $\beta=I/I_0$ is the moment of inertia of
the star in units of $I_0=MR_s^2$, $\kappa=\varepsilon\beta$, and
$\xi=\theta/\Theta$.

It is worth stressing that Eq.~(\ref{Mus_phi}) is the approximation
of a more complex expression involving Bessel functions (see, for
example, equations (50) and (51) of \citet{Muslimov1992}). Although
such an approximation makes Eq.~(\ref{Mus_phi}) more suitable to
describe the scalar potential at relatively large distances from the
star, after passing through the first zero of the zeroth-order
Bessel function, the scalar potential ceases to change noticeably
with the radial coordinate, tending to its asymptotic value. As a
result, in the rest of our analysis, when we consider the dependence
of the death-line on such parameters as the compactness of the star
$\varepsilon$ or the inclination angle $\chi$, we will use the
asymptotic expression provided by equation (\ref{Mus_phi}). With
this caveat in mind, the results we obtain become more and more
realistic in the far pair creation region and applicable in
principle if this region lies at distances larger than $R_p/\mu_1$
from the surface of the star (see also IS07), where $R_p$ is the
transverse size of the polar cap of the neutron star magnetosphere
and $\mu_1\approx2.4$ is the first root of the zeroth-order Bessel
function.

Expression (\ref{Mus_phi}) for the electromagnetic scalar potential
consists of two terms. The first one is purely general relativistic
and it contains the relativistic parameter $\kappa$ which tends to
zero in the case of a non-relativistic star. The second term also
contains some relativistic corrections, encoded in the functions
$H(\eta)$ and $f(\eta)$, but, unlike the first term, it does not
vanish in the case of a non-relativistic star. However, as noticed
by \citet{Muslimov1992}, unless the rotator is exactly orthogonal
($\chi=90^\circ$), the general relativistic component of the
electromagnetic scalar potential, and therefore of the accelerating
electric field in the polar cap region of the neutron star, is
approximately $10^2$ times greater than the scalar potential in the
flat-space limit (for typical values of the neutron star
parameters). Hence, in the model we are considering, the origin of
the electromagnetic scalar potential in the polar cap region of the
neutron star is mainly contributed by general relativistic
Lense-Thirring effect of dragging of inertial reference frames
through the parameter $\kappa$.

%--------------------------------------------------------
\subsection{Dependence of the death-line on the
  compactness parameter of the magnetar.}
\label{subsect11}

\begin{figure*}
\begin{center}
\includegraphics[width=0.6\textwidth]{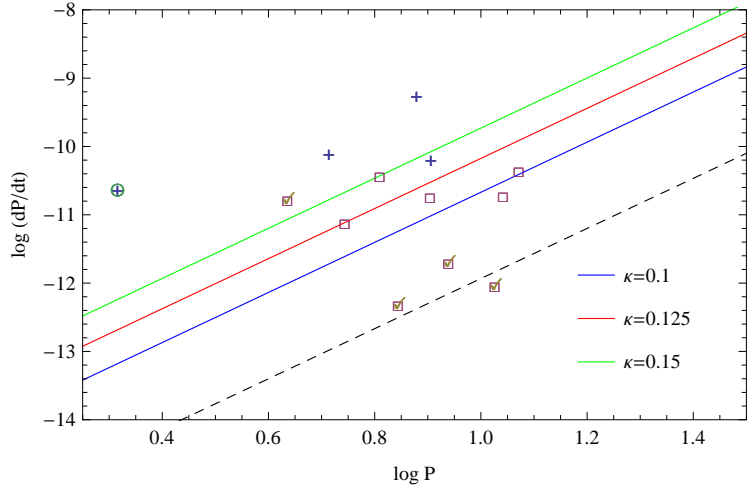}
\end{center}
\caption{ Death-lines for the aligned magnetar determined by
Eq.~(\ref{DLGR}) for different values of the parameter $\kappa$. The
dashed line indicates the position of the death-line from IS07. The
moment of inertia of the magnetar is taken to be $I=10^{45}
\mathrm{g}\ \mathrm{cm}^2$. Crosses and squares indicate the
position of soft gamma-ray repeaters and anomalous X-ray pulsars,
respectively. Anomalous X-ray pulsars from which the radio emission
has been registered are marked with ticks, radio-loud soft gamma-ray
repeater is enclosed in circle.} \label{Fig1}
\end{figure*}
We first consider the case $\chi=0$, namely when
the rotation axis of the neutron star is aligned with
the magnetic moment.
Then, after using
Eq.~(\ref{Mus_phi}) for the scalar potential in the polar cap
region of the magnetar, we obtain the following expression for
the minimum possible mean free path of the curvature photons in the
magnetar magnetosphere
\begin{equation}
\label{meanGR}
l_f=\frac{\rho}{\gamma_{max}}=\left(\frac{8}{3B_0}\frac{R_c^2}{R_s^3}\right)^3
\frac{R_s R_c f^3(1)}{\kappa^3}\xi^{-2}\left(1-\xi^2\right)^{-3}\ .
\end{equation}
In the derivation of the above equation we have used the expression for the
radius of curvature of the magnetic field line (see
IS07), namely $\rho=4\sqrt{zR}/3$,
written in cylindrical coordinates
$(\bar{r},z)$, where $\bar{r}$ denotes the distance from the dipolar
magnetic field axis to the point of observation, $z$ denotes the
distance from the stellar center to the point of observation along
the dipole axis and $R$ is the distance from the stellar center at
which a given field line will cross the $z=0$ plane when extended as
a dipole one, namely $R=z^3/r^2=const$ for a given field line.
Eq.~(\ref{meanGR}) differs from Eq.~(62) by IS07 in two
respects. Firstly,
we have adopted the parameter $\kappa$ rather than the
phenomenological parameter $(1-i_0)$, allowing us to explore the
dependence of the death-line position on the compactness parameter of
the magnetar in a rigorous way.
Secondly, our Eq.~(\ref{meanGR}) includes
the term $f^3(1)$ in the numerator, which represents
a genuine general relativistic correction.
The minimum of the mean free path $l_f$ as a function of the
coordinate $\xi$ is obtained for
$\xi=1/2$. Equating $l_{f\ min}$ to the stellar radius one
immediately obtains the value of the magnetic field for which the
generation of secondary plasma in the magnetosphere of the magnetar
is still possible:
\begin{equation}
B_0\gtrsim \left(\frac{\kappa}{f(1)}\right) \left(\frac{P}{1
\mathrm{s}}\right)^{7/3}\left(\frac{R_s}{10
\mathrm{km}}\right)^{-3}10^{12} \mathrm{G}\ ,
\end{equation}
which gives the expression for the death-line of the magnetars in the
form
\begin{equation}
\label{DLGR} \log\dot{P}=\frac{11}{3}\log
P-15.6-2\log\left(\frac{\kappa}{f(1)}\right)-6\log\left(\frac{R_s}{10
\mathrm{km}}\right)\ .
\end{equation}
There is unfortunately a great uncertainty in the determination of
both the moment of inertia and the mass of known magnetars, because
all of them are isolated objects not included in a binary system. In
the absence of more precise observational data one may, however, use
the approximate formula applicable for the neutron stars,
$\kappa=0.15 I_{45}/R_{6}$, derived by \citet{Muslimov1997} to
evaluate the parameter $\kappa$ for magnetars. Using the value
$I\sim10^{45} \mathrm{g}\ \mathrm{cm^2}$ \citep{Malheiro2011} we can
therefore draw the magnetar death-lines, reported in Fig.~\ref{Fig1},
for the different values of the radius of the magnetar. In addition,
we have also reported the observational data for magnetars as taken
from the ATNF catalog~\citep{Manchester2005}. The most relevant
result highlighted by Fig.~\ref{Fig1} is that {\em general relativistic
effects alone make the death-lines shift upwards}, and therefore move
magnetars in the radio-quiet zone below the death-line. This is in
agreement with the observations indicating that there is no periodic
radio emission from the magnetars in the range of frequencies
typical for ordinary pulsars. IS07 propose several arguments to
explain the fact that most of the magnetars in their Fig.1 are
placed in the radio-loud zone in the $P-\dot{P}$ diagram, contrary
to observations. On the other hand, such ad hoc arguments are not
necessary within our interpretation, where the radio-quietness is
naturally explained as due to purely relativistic
effects.
%The first reason suggested in IS07 is that there is
%a lack of synchrophotons in the magnetosphere of the magnetars since
%the first-generation particles are produced at the zeroth Landau
%level. This leads to the decreasing of the plasma generation
%multiplicity for the magnetars magnetosphere. The second argument is
%that the magnetars if not below the deathline are still fairly close
%to it and this may serve as a suppression factor for the secondary
%plasma generation.

It should be emphasized that according to our result the
non-relativistic case corresponds to zero $\kappa$, i.e. the
complete absence of the accelerating potential for the aligned
magnetar. This further clarifies the general relativistic origin of
the accelerating potential in the considered case, related to the
Lense-Thirring effect of the dragging of the inertial reference
frames.

\subsection{Dependence of the death-line on the
  inclination angle of the magnetar. }
\label{subsect12}

\begin{figure*}
\begin{center}\includegraphics[width=0.6\textwidth]{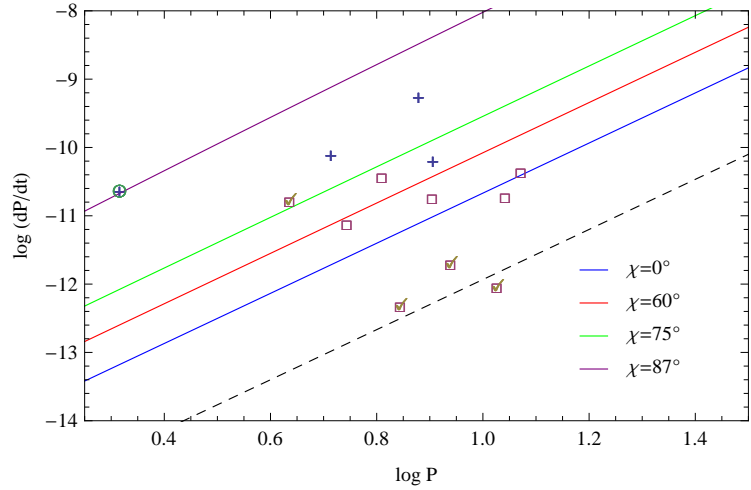}
\end{center}
\caption{ Death-lines for the misaligned magnetar for different
values of the inclination angle $\chi$. The value of $\kappa$ is
taken to be $0.1$. The dashed line indicates the position of the
death-line from IS07. The moment of inertia of the magnetar is taken
to be $I=10^{45} \mathrm{g} \mathrm{cm}^2$. Crosses and squares
indicate the position of soft gamma-ray repeaters and anomalous
X-ray pulsars, respectively.  Anomalous X-ray pulsars from which the
radio emission has been registered are marked with ticks, radio-loud
soft gamma-ray repeater is enclosed in circle.} \label{Fig2}
\end{figure*}

As a next step we investigate the dependence of the magnetar
death-line on the inclination angle between the angular
momentum of the magnetar and its magnetic moment.
By using Eq.~(\ref{Mus_phi})
with nonzero $\chi$, we obtain the following expression for the
minimum mean free path of the curvature photons

\begin{widetext}
\begin{equation}
l_{f\
min}=R_sR_c\left(\frac{8R_c^2}{3B_0R_s^3}\right)^3\xi_{min}^{-2}\left\{\left|\tilde{a}(1-\xi_{min}^2)+\tilde{b}\xi_{min}(1-\xi_{min}^2)\right|\right\}^{-3}\
,
\end{equation}
\end{widetext}

where
\begin{equation}
\tilde{a}=\frac{\kappa}{f(1)}\cos\chi\ , \quad
\tilde{b}=\frac{3}{4}\frac{R_s^{1/2}}{R_c^{1/2}}\frac{H(1)}{(f(1))^{3/2}}\left(\frac{\Theta(\eta)}{\Theta_0H(1)}-1\right)\sin\chi\
,
\end{equation}
while $\xi_{min}$ may be found as a solution of the cubic equation
\begin{equation}
\label{cubic}
11\tilde{b}\xi_{min}^3+8\tilde{a}\xi_{min}^2-5\tilde{b}\xi_{min}-2\tilde{b}\xi_{min}=0\
,
\end{equation}
obtained by setting to zero the derivative of the mean free path
$l_f$. We note that in the derivation of Eq.~(\ref{cubic}) we have
used the fact that $\tilde{b}>0$, so the minimum of the mean free
path $l_f$ corresponds to the value $\cos\phi=1$. As already
mentioned above, at the considered distances from the star surface
we are allowed to use the values of the functions $H(\eta)$ and
$f(\eta)$ in the limit of large $\eta$, therefore adopting
$H(\eta)\approx f(\eta)\approx 1$. The value of
$\Theta(\eta)/\Theta_0$, on the other hand, is determined after
using equation (\ref{theta}) in the small angle approximation. Under
these assumptions we have solved Eq.~(\ref{cubic}) numerically
finding the following expression for the death-line of the inclined
magnetar:

\begin{widetext}
\begin{equation}
B>2^{-\frac{8}{3}}3\xi_{min}^{-\frac{2}{3}}\left\{\left|\frac{\kappa}{f(1)}\cos\chi(1-\xi_{min}^2)
+\frac{3}{4}\frac{1}{(f(1))^{3/2}}\sqrt{\frac{R_s}{R_c}}\left(\frac{\Theta(\eta)}{\Theta_0}-H(1)\right)\sin\chi\right|\right\}^{-1}\left(\frac{P}{1
s}\right)^{\frac{7}{3}}\left(\frac{R_s}{10 km}\right)^{-3}10^{12}
\mathrm{G}\ .
\end{equation}
\end{widetext}

In Fig.\ref{Fig2} we have reported a few death-lines for magnetars
having different inclination angles, while keeping the same
compactness parameter $\kappa=0.1$. This figure shows that, by
increasing the angle $\chi$ between the angular momentum vector of
the star and its magnetic moment, the death-line is shifted upwards.
Although this effect is even more pronounced than that produced
by purely general relativistic effects, its physical relevance for
explaining the radio-quietness of magnetars cannot be
over-emphasized, since it would imply a implausible strong
misalignment in all magnetars that have been observed.

\section{Death-line for the rotating and oscillating magnetars}
\label{deathline}

In this section we consider magnetars that are subject
to toroidal oscillations, relying on results obtained by
\cite{Timokhin2007} and \cite{Morozova2010}.
In spherical coordinates $(r,\theta,\phi)$
the velocity field of an oscillating neutron star can be
written as~\citep{Unno1989}
\begin{equation}
\delta
v^{\hat{i}}=\left\{0,\frac{1}{\sin\theta}\partial_{\phi}Y_{lm}(\theta,\phi),
-\partial_{\theta}Y_{lm}(\theta,\phi)\right\}\tilde{\eta}(r)e^{-i\omega
t}\ ,
\end{equation}
where $\tilde{\eta}$ is the radial eigenfunction
expressing the amplitude of the
oscillation, $\omega$ is the frequency of oscillation and the
orthonormal functions $Y_{lm}(\theta,\phi)$ are the
eigenfunctions of the Laplacian in spherical coordinates.

The electromagnetic scalar potential in the polar cap region of
rotating and oscillating aligned neutron star magnetosphere has been
computed by \cite{Morozova2010} and it is given by

\begin{widetext}
\begin{equation}
\label{Psi1}
\Psi(\theta,\phi)=\frac{B_0}{2}\frac{R_s^3}{R_c^2}\frac{\kappa}{f(1)}\left(1-\xi^2\right)-
e^{-i\omega
t}\tilde{\eta}(R_s)B_0R_s\sum_{l=0}^{\infty}\sum_{m=-l}^{l}Y_{lm}(\theta,\phi)\
.
\end{equation}
\end{widetext}

This equation differs from the analogous equation (33) by
\cite{Morozova2010} in one fundamental respect. Namely, the
oscillatory part of the scalar potential is missing the term
proportional to $1/r^2$ that was reported in Eq.~(33) by
\cite{Morozova2010}, because we have taken the asymptotic value as
discussed above.

For any particular mode $(l,m)$, we use
the approximation $Y_{lm}(\theta,\phi)\approx
A_{lm}(\phi)\theta^m$, valid in the limit of small polar angles
$\theta$, where the terms $A_{lm}(\phi)$ have
real parts given by
\begin{eqnarray}
\label{listA}
A_{00}&=&\frac{1}{\sqrt{4\pi}}\ , \\
A_{10}&=&\sqrt{\frac{3}{4\pi}} \ , \\
A_{11}&=&-\sqrt{\frac{3}{8\pi}}\cos\phi \ , \\
A_{20}&=&\sqrt{\frac{5}{4\pi}} \ , \\
A_{21}&=&-3\sqrt{\frac{5}{24\pi}}\cos\phi\ .
\end{eqnarray}
In addition, since we do not explore the time evolution of the
system, in the calculations that follow we drop the time dependence
of the oscillatory term (computed at time $t=0$ as in
\cite{Morozova2010}), we then recall the definition
$\xi=\theta/\Theta$, and we therefore rewrite Eq.~(\ref{Psi1}) as
\begin{eqnarray}
\label{Psi2}
\Psi(\xi,\phi)_{lm}&=&\frac{B_0}{2}\frac{R_s^3}{R_c^2}\frac{\kappa}{f(1)}\left(1-\xi^2\right)-\nonumber \\
&&\tilde{\eta}(R_s)\frac{B_0}{f^m(1)}\frac{R_s^{\frac{m}{2}+1}}{R_c^{\frac{m}{2}}}\xi^{m}A_{lm}(\phi)\
,
\end{eqnarray}
where one takes into account the expressions (\ref{theta}) and
(\ref{theta0}) for the $\Theta$  and $\Theta_0$, respectively.
\begin{figure*}
\begin{center}\includegraphics[width=0.48\textwidth]{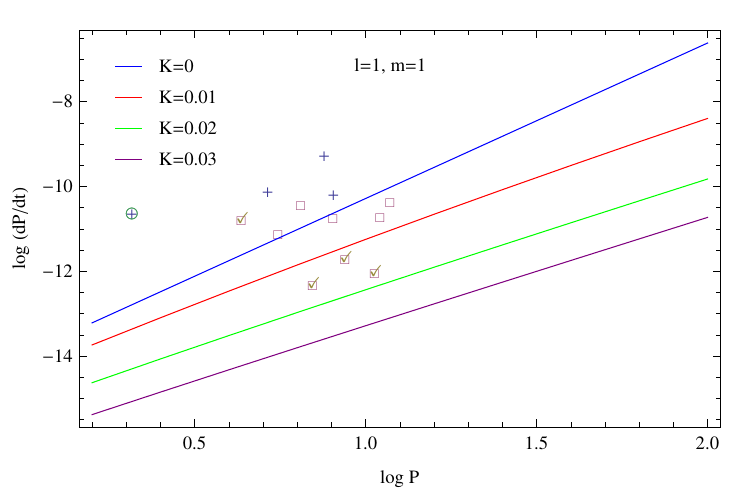}
\includegraphics[width=0.48\textwidth]{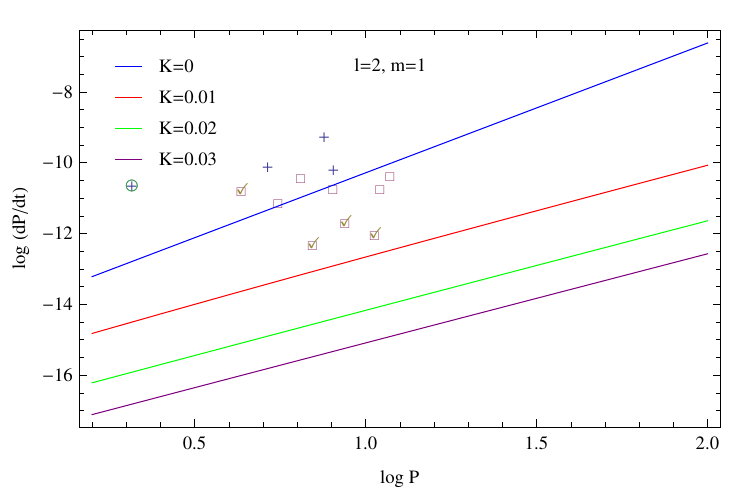}\end{center}
\caption{Death-lines for rotating and oscillating magnetars in the
$P-\dot{P}$ diagram. The left panel corresponds to the mode $(1,1)$
and values of $K=0\ ,0.01\ ,0.02\ ,0.03$. The right panel
corresponds to the mode $(2,1)$ and values of $K=0\ ,0.01\ ,0.02\
,0.03$. Other parameters are taken to be $R_s=10 \mathrm{km}$,
$M=2M_{\bigodot}$ and $\kappa=0.15$. Crosses and squares indicate
the position of soft gamma-ray repeaters and anomalous X-ray
pulsars, respectively. Anomalous X-ray pulsars from which the radio
emission has been registered are marked with ticks, radio-loud soft
gamma-ray repeater is enclosed in circle.} \label{Fig3}
\end{figure*}

At this point, by repeating the procedure already described in
Sec.~\ref{subsect11} and \ref{subsect12}, but with the scalar
potential in the polar cap region given by Eq.~(\ref{Psi2}), we find
the  mean free path of the curvature photon in the form

\begin{widetext}
\begin{equation}
\label{l_f_oscill} l_f=R_sR_c\left(\frac{4}{3}\right)^{3}\xi^{-2}
\left\{\left|\frac{B_0}{2}\frac{R_s^3}{R_c^2}\frac{\kappa}{f(1)}\left(1-\xi^2\right)-
\tilde{\eta}(R_s)\frac{B_0}{f^m(1)}\frac{R_s^{\frac{m}{2}+1}}{R_c^{\frac{m}{2}}}\xi^{m}A_{lm}(\phi)\right|\right\}^{-3}\
.
\end{equation}
\end{widetext}

The minima of \eqref{l_f_oscill} clearly depends on
the choice of the mode $(l,m)$ that is considered. For
example, for $m\neq0$, the minimum of $l_f$ comes after
solving the following quadratic equation
\begin{equation}
\label{minimum} 8a\xi_{min}^2+(2b+3mb)\xi_{min}^m=2a\ ,
\end{equation}
where
\begin{equation}
a=\frac{B_0}{2}\frac{R_s^3}{R_c^2}\frac{\kappa}{f(1)}\ ,\qquad
b=\tilde{\eta}(R_s)\frac{B_0}{f^m(1)}\frac{R_s^{\frac{m}{2}+1}}{R_c^{\frac{m}{2}}}A_{lm}(\phi)\
.
\end{equation}
When $m=1$ and $m=2$ the solution of equation (\ref{minimum}) is
given by
\begin{eqnarray}
\xi_{min}&=&\frac{-5b+\sqrt{25b^2+64a^2}}{16a}
\ \ \ \ \ \ \ \ \ \rm{for} \ \ \ \ m=1 \ , and \\
\xi_{min}&=&\frac{1}{2}\sqrt{\frac{a}{a+b}} \hspace{1.1cm} \ \ \ \ \ \ \ \ \ \ \ \ \ \ \ \ \rm{for}
\ \ \ \ m=2 \ .
\end{eqnarray}
In the absence of oscillations, namely when $b=0$, both
cases $m=1$ and $m=2$ give the
minimum of $l_f$ at the value $\xi_{min}=1/2$, corresponding to the
case of pure rotation~\citep{Istomin2007}.
After performing an averaging on the azimuthal
angle $\phi$, we derive the condition for radio emission
on the intensity of
the magnetic field, which, for $m\neq 0$, is given by

\begin{widetext}
\begin{eqnarray}
&& B>2^{-\frac{8}{3}}6\pi\left\{\int_0^{2\pi}\xi_{min}^{2/3}
\left|\frac{\kappa}{f(1)}(1-\xi_{min}^2)-2\frac{\tilde{\eta}(R_s)}{f^m(1)}\left(\frac{R_s}{R_c}\right)^{\frac{m}{2}-2}\xi_{min}^m
A_{lm}(\phi)\right|d\phi\right\}^{-1}
 \times\left(\frac{P}{1
s}\right)^{\frac{7}{3}}\left(\frac{R_s}{10 km}\right)^{-3}10^{12}
\mathrm{G}\ .\nonumber \\
&&
\end{eqnarray}
\end{widetext}

We then proceed in complete analogy with
Sec.~\ref{subsect11}, namely by
assuming that the magnetar
spins down due to the
generation of magnetodipole radiation, so that
$B_0\simeq2(P\dot{P}_{-15})^{1/2}10^{12} \mathrm{G}$. In
this way the equation of the death-lines for oscillating
and rotating neutron stars is

\begin{widetext}
\begin{eqnarray}
&&\log \dot{P}_{-15}=\frac{11}{3}\log P-0.6+\log(C^2)\ , \\
&&C=2^{-\frac{8}{3}}6\pi\left\{\int_0^{2\pi}\xi_{min}^{2/3}
\left|\frac{\kappa}{f(1)}(1-\xi_{min}^2)-2\frac{\tilde{\eta}(R_s)}{f^m(1)}\left(\frac{R_s}{R_c}\right)^{\frac{m}{2}-2}\xi_{min}^m
A_{lm}(\phi)\right|d\phi\right\}^{-1}\ ,
\end{eqnarray}
\end{widetext}

where $\dot{P}_{-15}$ is the period time derivative measured in
$10^{-15} \mathrm{s\ s^{-1}}$. The amplitude of the oscillation is
now parametrized in terms of the small number
$K=\tilde{\eta}(1)/\Omega R$, giving the ratio between the velocity
of oscillations and the linear rotational velocity of magnetar.
Fig.~\ref{Fig3} reports the death-lines for rotating as well as
oscillating magnetars for two modes of oscillations and different
values of the parameter $K$. In particular, the left panel shows the
case $l=1,m=1$, while the right panel shows the case $l=2,m=1$, thus
providing the two most relevant physical cases for non-axisymmetric
modes.
Unlike the effects induced by general relativistic corrections and
by the inclination angle $\chi$, which only produce shifting of the
death-lines without affecting their slope, changes in the intensity
of the oscillation affect both the slope and the shifting of the
death-lines. In particular, larger values of $K$ produce smaller
slopes (though converging to a constant value) and downward shifting
in the $P-\dot{P}$ diagram. On the overall, stronger oscillations of
the modes with $m=1$ contribute to radio-loudness of magnetars. This
result is in agreement with the work by \cite{Morozova2010}, where
it was found that that oscillation modes with $m=1$ considerably
increase the electromagnetic energy losses from the polar cap region
of the neutron star, which may be several times larger than in the
case when no oscillations are present.

In a recent paper, \cite{Timokhin2008} invoked stellar oscillations
to explain the observed quasi-periodic oscillations in the tail of
soft gamma repeater giant flares. Such oscillations lie in the range
between 18 Hz and 1800 Hz, and are widely interpreted as shear modes
of the solid crust of the neutron stars~\citep{Steiner2009}.
\cite{Timokhin2008} showed that the stellar oscillations should be
of the order of $1\%$ of the stellar radii, in order to explain the
observed phenomenology. It is interesting to note that, when
translated in terms of our parameter $K$, these estimates imply
$K\sim 0.1$ or higher, while we have shown that noticeable effects
on the magnetospheric features of the magnetars are already present
for  $K\sim0.01-0.03$.

\begin{figure*}
\begin{center}\includegraphics[width=0.8\textwidth]{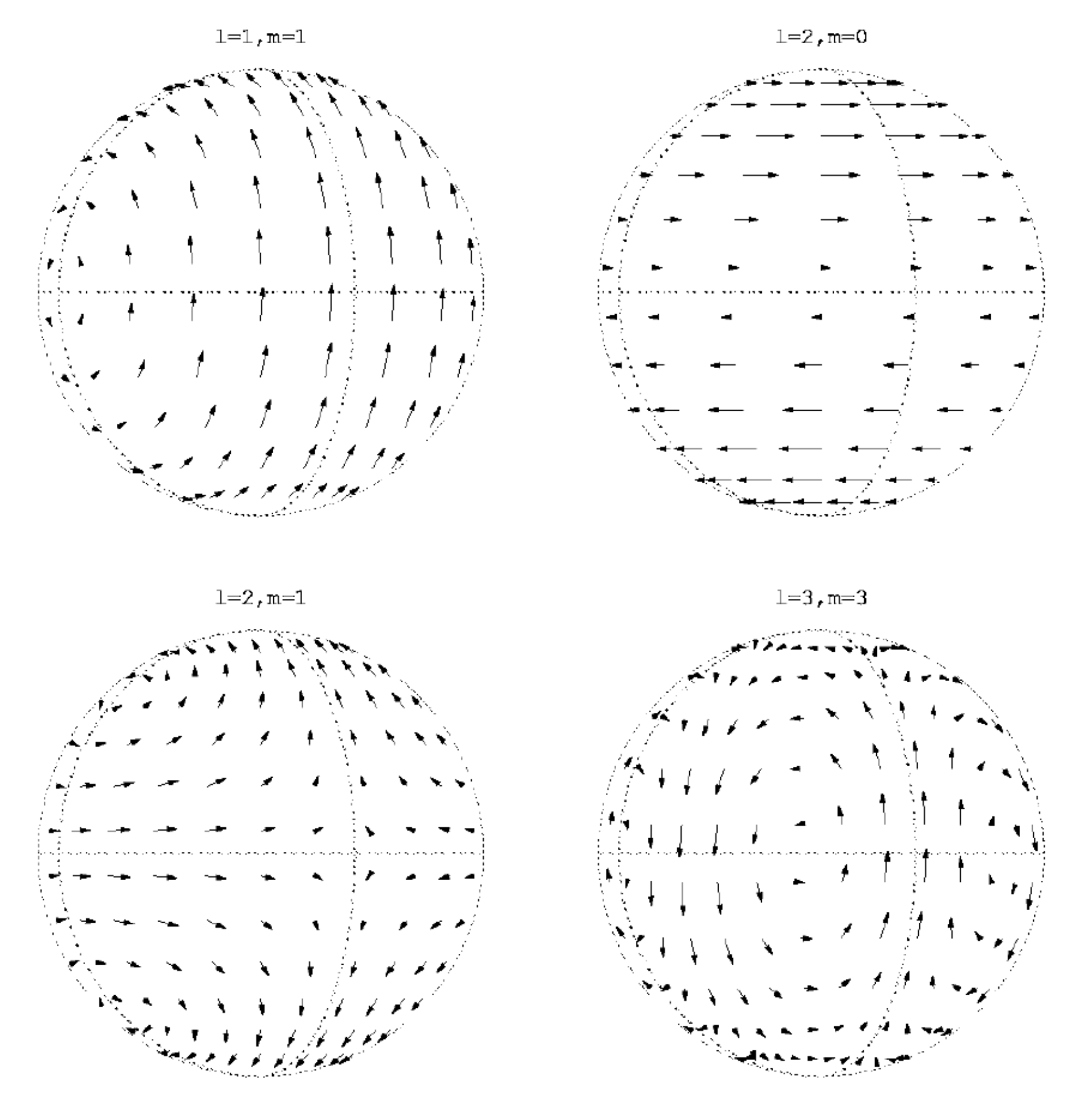}\end{center}
\caption{Velocity distribution for a few different
modes of stellar oscillations.} \label{Fig4}
\end{figure*}

A complementary information to that of Fig. \ref{Fig3} is provided
by Fig. \ref{Fig4}, where we have plotted the velocity field for a
few modes of stellar oscillations (see also \cite{Timokhin2000}).
The electron-positron plasma in the neutron star magnetosphere is
expected to be continuously generated in the polar cap region, where
magnetic field lines are open and the plasma may freely escape from
the surface of the star to infinity. When the star rotates, the
linear velocity of the stellar surface motion is of course
proportional to the distance from the axis of rotation to the
considered point. Therefore, in the polar cap region this velocity
is proportional to the polar angle. When we limit our attention to
the polar cap region, namely by looking at the top of each sphere
reported in Fig.~\ref{Fig4}, we see that the velocity distribution
for the oscillatory modes with $m=0$ has the same form as for the
case of pure rotation. For the modes $m=1$, on the other hand, the
velocity of oscillations remains almost constant across the polar
cap region. This further indicates that the
  modes with $m=1$ have greater impact on the
  magnetosphere processes and, therefore, merit more attention.

\section{Conclusions}
\label{conclusions}

It was not until 2006 that the first detection of a radio magnetar
was reported. To date, only two radio magnetars have been confirmed,
namely XTE J1810|197 and 1E 1547.0-5408 and it remains unclear what
is the physical mechanism preventing the radio-loudness of these
sources.

In this work, by using the general relativistic expression for the
accelerating electromagnetic scalar potential in the vicinity of the
polar cap, we have performed a detailed analysis of the position of
the death-line in the $P-\dot{P}$ diagram. Our results can be
summarized as follows:
\begin{itemize}
\item When the compactness of the neutron star is
  increased, the death line shifts
  upwards in the $P-\dot{P}$ diagram, pushing the
  magnetar in the radio-quiet region. This is a purely
  general relativistic effect.
\item When the inclination angle $\chi$ between the
  angular momentum vector of the neutron star and its
  magnetic moment is increased, the death-line shifts
  upwards in the $P-\dot{P}$ diagram, pushing the
  magnetar in the radioquiet region.
\item On the contrary, when oscillations of the magnetar are taken into
  account, the radio emission from
  the magnetosphere is generally favored.
In fact, the major effect of oscillations is to amplify the scalar
potential in the polar cap region of the magnetar magnetosphere. As
a result, the energy of primary particles that are pulled out and
accelerated in the vicinity of the stellar surface is enhanced, and
the probability of effective electron-positron plasma generation is
increased. The largest effect is expected for oscillation
modes with $m=1$, whose
  velocity field is almost uniform in the polar cap
region.

\end{itemize}

It is worth stressing that, according to
  our explanation, there is not a unique death line which
  is valid for the entire class of magnetars. On the contrary, each source
  has its own death line that is determined by individual
  physical conditions.
 This also implies that
 a source in the $P-\dot{P}$ diagram may be radio-quiet
 while another source, located {\em below} the first one,
 is radio-loud,
 just because their death-lines are different.
While providing an effective explanation of magnetar
radio-quieteness in terms of the stellar compactness and of the
misalingment between angular momentum and magnetic moment, our
results may also indicate that the unusual radio emission observed
from some magnetars may be related to the generation of oscillations
in the magnetar crust. This hypothesis is sustained by the
registration of preceding bursts from these magnetars. According to
this interpretation, one may expect that after the magnetar burst,
which produces mechanical oscillations of the stellar crust, the
conditions for radio emission within the magnetospehere are
satisfied. However, because of the inevitable damping of the
oscillations, after a damping time the magnetar will turn back to
the radioquiet region in the $P-\dot{P}$ diagram (the deathline will
shift back upwards). This is qualitatively in agreement with
observations showing that in radio-loud magnetars the radio emission
reveals strong fluctuations in time. This is particularly true for
the candidate radio-magnetars which may be located at the border
between radio-loud and radio-quiet regions (see \cite{Malofeev2007},
\cite{Malofeev2010}). Future more accurate observations are likely
to verify or refute our interpretation by testing if there is a
close connection between the burst activity of magnetars and the
generation of the radio emission in the magnetar magnetosphere.

%%%%%%%%%%%%%%%%%%%%%%%%%%%%%%%%%%%%%%%%%%%%%%%%%%%%%%%%%%%%%%%%%%%%%%%%%
\section*{Acknowledgments}
%%%%%%%%%%%%%%%%%%%%%%%%%%%%%%%%%%%%%%%%%%%%%%%%%%%%%%%%%%%%%%%%%%%%%%%%%

This research is supported in part by Projects No. FA-F2-079 and No.
FA-F2-F061 of the UzAS and by the ICTP grant PRJ-29. Authors would
like to acknowledge the hospitality of the Albert Einstein Institute
for Gravitational Physics, Golm, where the most part of the research
has been performed.

\bibliographystyle{mn2e}

%\bibliography{Mag_DL_Osc}
%\bibliography{aeireferences}

\label{lastpage}

\end{document}